\documentstyle[preprint,epsf,aps]{revtex}
\draft

\begin{document}
\title{Inhibition of Decoherence due to Decay in a Continuum}
\author{G.S. Agarwal$^{1,2}$, M.O. Scully $^{2,3}$ and H. Walther
$^2$}
\address{1. Physical Research Laboratory, Navrangpura, Ahmedabad-
380009, India\\2. Max-Planck-Institut F\"{u}r Quantenoptik, D-85748 Garching, 
Germany.\\3. Texas A\&M, Department of Physics, College Station,
TX77843-4242, USA}
\date{\today}

\maketitle
\begin{abstract}
We propose a scheme for slowing down decay into a continuum. We make
use of a sequence of ultrashort $2\pi$-pulses applied on an auxiliary
transition of the system so that there is a destructive interference
between the two transition amplitudes - one before the application
of the pulse and the other after the application of the pulse. We
give explicit results for a structured continuum. Our scheme can 
also inhibit unwanted transitions.\\
\pacs {PACS Number(s):~  03.65. Bz, ~  03.67 - a, ~  42.50. Md}
\end{abstract}
\newpage
One of the fundamental causes of noise at the optical frequencies 
is the intrinsic spontaneous emission. It is well known that the
performance of many systems is limited by spontaneous emission.
For example the noise figure of an amplifier is determined by 
spontaneous emission. One of the challenges therefore is to 
find ways in which spontaneous emission noise can be reduced if
not totally inhibited. Several proposals exist in literature 
for reducing spontaneous emission noise. These include - placing of atoms in a photonic bandgap structure 
\cite{yablonovitch,quang}, use of external fields \cite{lange}
and quantum interferences \cite{saran,zhu}. As is well known, 
spontaneous emission noise arises from considerations based on the interaction of an atom
with the vacuum of the electromagnetic field. The vacuum acts like a zero 
temperature bath. It may be noted that the question of 
inhibiting the effects of a heat bath is being extensively 
studied and several new proposals exist for such an inhibition
\cite{viola,vitali,tombesi,gsa,brune}. For example, for a spin 
system interacting with a bath the decoherence can be slowed
\cite{viola} by applying a sequence of $\pi$ 
pulses applied at invervals of a short period $\tau$ which is less
than the bath correlation time.

In this letter we consider the case of spontaneous emission from an excited atom.
One has considerably more freedom with atoms than with spins as in
the former case we could use a different transition to control
spontaneous emission. We propose a scheme to suppress spontaneous
emission on (say) the emission from the state $|e\rangle$ to $|g\rangle$ by using an auxiliary
transition $|g\rangle$ 
to $|f\rangle$. We apply a sequence of ultrashort $2\pi$-pulses 
separated by an interval $\tau$. We demonstrate how a 
destructive interference between the evolution from $t_o$ to
$t_o+\tau$ and $t_o+\tau$ to $t_o+2\tau$ can lead to 
suppression of decoherence. It should be borne in mind that the
decoherence time scale is generally proportional to the decay time.
The destructive interference is related to the very remarkable 
property that after the application of a $2\pi$ pulse the state 
$|g\rangle$ acquires a
phase shift of $\pi$. An important by-product of our 
investigation is the possibility of suppressing undesirable
weak transitions. 

We first consider a simple case which is adequate to describe 
the main idea and which by itself is very relevant to the subject of 
quantum computation\cite{plenio}. Consider a quantum system which can make an 
{\it unwanted} weak transition from the state $|g\rangle$ to
$|e\rangle$ as a result of a perturbation $v$.
Let $\delta$ be the detuning i.e. the perturbation need not be resonant
with the transition $|g\rangle\leftrightarrow|e\rangle$. The interaction
Hamiltonian in the interaction picture is
\begin{equation}
H_1(t) = \hbar v|e\rangle\langle g|e^{-i\delta t} +{\rm H.c}.
\end{equation}
The perturbation theory leads to the probability of transition
\begin{equation}
p_{eg}=|v|^2 \frac{\sin^2(\delta t/2)}{(\delta/2)^2} .
\end{equation}
We next demonstrate how this unwanted transition could be inhibited
by applying a sequence of very short $2\pi$-pulses \cite{scully}
on the transition
$|g\rangle\leftrightarrow |l\rangle$ [Fig.1]. We thus divide the total
time interval into a large number $2N$  of short intervals $\tau$.The
system evolves under $v$ from $t_o$ to $t_o+\tau$. At $t_o+\tau$
we apply an ultra-short $2\pi$-pulse on the transition
$|g\rangle\leftrightarrow|l\rangle$. The system evolves from $t_o+
\tau$ to $t_o+2\tau$ under $v$. At the instant $t_o+2\tau$ 
the $2\pi$-pulse is applied again. This process is repeated $N$
number of times. The system then evolves as follows:- For~ 
$t_o ~< t~<~ t_o+\tau,$ we have  
\begin{equation}
|\psi (t)\rangle\sim |g\rangle-iv|e\rangle\left(X(t-t_o)/(-i\delta)
\right)e^{-i\delta t_o};X(t)=(e^{-i\delta t}-1).
\end{equation} 
At time $t_o+\tau$ after the application of $2\pi$-pulse, the state 
of the system is
\begin{equation}
|\psi\rangle=-|g\rangle -iv|e\rangle X (\tau)e^{-i\delta t_o}/
(-i\delta).
\end{equation}
At a time $t_o+2\tau$ just before the application of the second
pulse the state would evolve into
\begin{equation}
|\psi\rangle = -|g\rangle-iv|e\rangle(X(\tau)/(-i\delta)) e^{-i\delta t_o}
+iv|e\rangle (X(\tau)/(-i\delta))e^{-i\delta (t_o+\tau)}
~+~0(v^2),
\end{equation}
which on application of the second $2\pi$-pulse changes to 
\begin{equation}
|\psi\rangle\equiv|g\rangle+iv|e\rangle(X(\tau))^2 
e^{-i\delta t_o}/(-i\delta)~ +~ 0(v^2).
\end{equation}
The transition amplitude $\chi_{eg}$ at the end of one cycle consisting
of evolution of the system from $t_o$ to $t_o+2\tau$ but with $2\pi$
-pulses applied at $t_o+\tau$ and $t_o+2\tau$ will be 

\begin{equation}
\chi_{eg} = iv(X(\tau))^2 e^{-i\delta t_o}/(-i\delta)+ 0 (v^2).
\end{equation}
The transition amplitude at the end of $N$ such cycles will be 
\begin{equation}
\chi_{eg}= iv\frac{(X(\tau))^2}{(-i\delta)}\sum_{p=0}^{N-1} e^{-2i\delta
\tau p-i\delta t_o} ,
\end{equation}
which leads to the following result for the net transition probability
\begin{equation}
\tilde{p}_{eg} = |v|^2 \tan^2\left(\frac{\delta\tau}{2}\right)
 \frac{\sin^2\left(\frac{\delta}{2}(2\tau N)\right)}{\left({\delta/2}
 \right)^2}.
\end{equation}
On using (2) we have one of our key results
\begin{equation}
\tilde{p}_{eg} = \tan^2\left(\frac{\delta\tau}{2}\right)p_{eg}.
\end{equation}
We have thus proved that the application of a sequence of $2\pi$-pulses
on an auxiliary transition leads to the suppression of an unwanted 
transition provided that the small interval and the  detuning 
$\delta$ are chosen such that
\begin{equation}
\tan^2 (\frac{\delta\tau}{2})~ \ll~ 1 .
\end{equation}
The suppression arises from a {\it destructive interference} of the transition
amplitudes (second and third terms in Eq.(5)). This distructive 
interference is due to a phase change of the state $|g\rangle$ (and
\underline{not} of $|e\rangle$) by $\pi$ due to the application of 
the $2\pi$-pulse. This also explains our choice of an auxiliary transition
for the application of the $2\pi$-pulse as we selectively want to 
produce a phase change so that the interference can occur. 
One might think that the procedure we describe is just the 
quantum Zeno effect \cite{misra}. We emphasize that it is {\em not} as we do not carry out
repeated measurements. However it can be considered at best an analog
of Zeno effect in the spirit of \cite{heinzen}, though collapse of the 
state is considered to be an essential ingredient for quantum Zeno
effect \cite{namiki}.

We next demonstrate how the above procedure can be used to possibly
inhibit spontaneous emission. We show the procedure schematically
in Fig 2. The atom makes a transition from the excited state
$|e\rangle$ to the ground state $|g\rangle$ by emitting
a photon. The photon can be emitted in any mode $\omega_k$ of the 
vacuum of the radiation field. The polarization of the emitted photon
will be determined by the direction of the dipole matrix element. The
interaction Hamiltonian in the interaction picture is \cite{saran}.
\begin{eqnarray}
H_1(t)&=&\hbar\sum_k |e\rangle\langle g|g_k a_k e^{-i\delta_k t}+
{\rm H.c};\\ \nonumber
\delta_k&=&\omega_k-\omega_{eg}.
\end{eqnarray}
For brevity, we do not display the polarization $``s"$ and vectorial indices
of the mode. Thus $k$ really stands for $(k,s)$. In (12), $a_k$ is the
annihilation operator for the mode $k$ of the radiation field. We now
follow the precedure leading to Eq.(9). We quote the results of calculations
for the transition probabilities with $(\tilde{p}_{ge})$ and without 
$(p_{ge})$ the application of $2\pi$-pulses
\begin{equation}
\tilde{p}_{ge} =\sum_k |g_k|^2 \tan^2 \left(\frac{\delta_k\tau}{2}\right)
\frac{\sin^2\left(\frac{\delta k}{2} 2\tau N\right)}{\left
(\delta_k/2\right)^2},
\end{equation}
\begin{equation}
p_{ge}=\sum_k |g_k|^2~ \frac{\sin^2\left(\frac{\delta k}{2}2\tau N
\right)}
{\left(\delta_k/2\right)^2}.
\end{equation}
We first note how (14) leads to the standard result and how the 
Einstien A-coefficient emerges.
If $t=2N\tau$, then 
\begin{equation}
\partial p_{ge}/\partial t=\sum_k 2|g_k|^2~ \frac{\sin \delta_k t}
{\delta_k}
\end{equation}
which, under the assumption that the observation time $t$ is large 
compared to the width of $(\delta_k)$ values (which is of the order
of $\omega_{eg}$ for spontaneous emission in free space),  reduces to
the standard expression for the Einstein A-coefficient
\begin{equation}
\partial p_{ge}/\partial t = 2\pi\sum |g_k|^2\delta~(\omega_k-
\omega_{eg})\equiv A.
\end{equation}
We next examine the conditions under which the presence of the quantum
interference term $\tan^2\left(\delta_k\tau/2\right)$
in Eq. (13) can lead to the suppression of spontaneous emission. 
Let us consider a kind of one dimensional model in which
we can replace (13) by
\begin{equation}
\tilde{p}_{ge}\equiv\int_{-\omega_{eg}}^{\infty}dx~\tan^2\left(\frac
{x\tau}{2}\right)~\frac{\sin^2(\frac{x}{2} 2\tau N)}{(x/2)^2}
\rho(x),
\end{equation}
where $\rho(x)$ is the density of states for the one dimensional
vacuum. For $\rho(x),$ we can choose any of the functions like
the Lorentzian or the exponential
\begin{equation}
\rho(x) = \frac{\rho_o\Gamma/\pi}{(x^2+\Gamma^2)}~~or ,~~~ 
\rho(x)=\frac{\rho_o}{2\Gamma}~e^{-|x|/\Gamma}
\end{equation}
with $\Gamma\ll\omega_{eg}$ and where $\rho_o$ is related to the 
square of the dipole matrix element. Note that these choices correspond 
to what are known as the ``structured" vacuua\cite{bay}. Thus we are essentially
asking the question - under what conditions the emission into 
structured vacuua can be inhibited. We use the exponential model in
(17) and show a typical result in Fig.3. For comparison we also
show the result $p_{ge}$. On comparing the two results we see that 
the use of a sequence of $2\pi$-pulses applied in the manner shown
in Fig.2 can suppress to a large extent, the decay into a 
structured continuum.

We next consider briefly the question of spontaneous emission in
free space. We convert (13) into an integral using the standard 
expressions for $g_k$ and by letting the quantization volume go to 
infinity. The transition probabity $\tilde{p}_{ge}$ depends on the 
following integral
\begin{eqnarray}
I &=& \int_{-1}^{\infty} dx(x+1)^3~ \tan^2 \left(\frac{x\tau}{2}\right)
~ \frac{\sin^2(\frac{x}{2} 2\tau N)}{{(x/2)}^2},\\ \nonumber
x &=& (\omega-\omega_{eg})/\omega_{eg},~ \tau\rightarrow\omega_{eg}
\tau.
\end{eqnarray}
Note that the sine function is sharply peaked at $x=0$ if $2N\tau$
is large. This is what enabled us to simplify the expression for 
$p_{ge}$ leading to (16). However, we now have the interference factor
$\tan^2\left(x\tau/2\right)$ which starts growing as 
$x\tau/2\rightarrow\pi/2$. Thus the idea of $2\pi$- 
pulse induced slowing down of decoherence will obviously work if
the bath has a cut-off such that $\tan^2(x\tau/2)$ remains much 
smaller than unity. In order to examine further the 
question of slowing down of the decoherence we analyse the 
integrand $\tilde{I}$ in (19). We can rewrite the integrand in the
form
\begin{eqnarray}
\tilde{I}\equiv(x+1)^3~ \frac{16 \sin^4 (x\tau/2)}{x^2}
\left [ \frac{\sin^2(x \tau N)}{\sin^2(x\tau)}\right ].
\end{eqnarray}
Note that the integrand oscillates very rapidly due to the term
$\sin^2(x\tau N).$ Though it has no singularities as the term in the 
square bracket is well behaved, it still grows as $N^2$ whenever
$x\tau = n\pi$. Such a growth problem can be avoided if the vacuum
mode frequencies are such that $x\tau\ll\pi$. If $x$ is chosen to
be of the order of unity, then the pulses have to be applied at 
{\it intervals much smaller than} $\omega_{eg}^{-1}$, which is a 
difficult task in the optical domain though {\it not} for 
{\it Rydberg transitions.} Note that with pulses applied at intervals less than 
$\omega_{eg}^{-1}$, the details of spontaneous emission become sensitive to the 
structure of the bath as shown in the Fig.4. We note that Vitali and Tombesi \cite
{vitali} have examined the damping of a harmonic oscillator 
interacting with a broad band bath and have reached similar
conclusions. The nature of the integrand in (19) also suggests that 
if the interval $\tau > \omega_{eg}^{-1}$,  then we have the possibility
of accelerating spontaneous emission (Fig.4; curves for 
$\omega_{eg}\tau=\pi)$ which is suggestive of an 
effect that is an analog of recently discovered quantum anti-Zeno effect
\cite{kofman}.

In conclusion,  we have shown how a sequence of $2\pi$-pulses applied
on an auxiliary transition in a system can slow down considerably
the decay into a continuum \cite{taylor}. The same scheme also enables one to
inhibit unwanted transitions.

The authors thank K. Kapale and  S. Menon for preparing the 
figures.

\begin{figure}
\epsfxsize 3 in
\centerline{\epsfbox{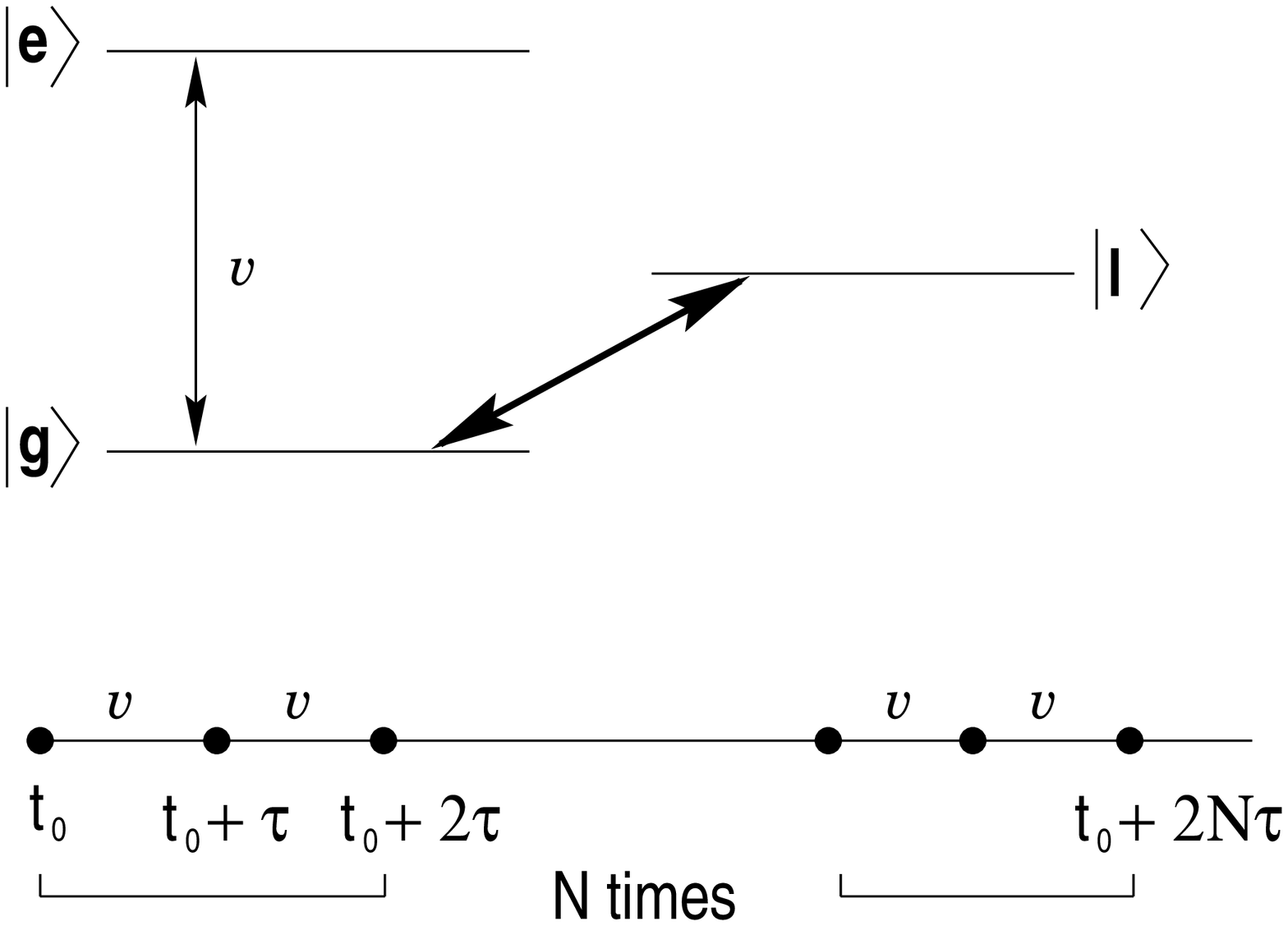}}
\caption{Scheme for suppression of an unwanted transition 
$|g\rangle\rightarrow|e\rangle$ caused by a weak off-resonant
perturbation $v$. The bold arrow represents the $2\pi$-pulse
on the transition $|g\rangle\leftrightarrow|l\rangle$. The crosses
denote the times, when an ultrashort $2\pi$-pulse is applied. In
between the pulses the system evolves under $v$.}
\end{figure}
\newpage
\begin{figure}
\epsfxsize 3 in
\centerline{\epsfbox{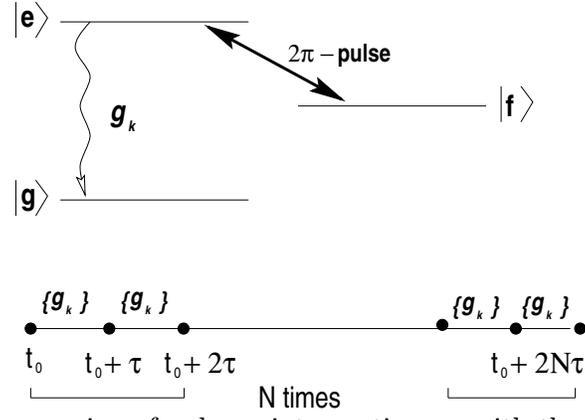}}
\caption{Scheme for suppression of a decay into continuum with 
the atom going from $|e\rangle$ to $|g\rangle$ by the emission
of a photon, other specification same as in Fig.1.}
\end{figure}
\newpage
\begin{figure}
\epsfxsize 3 in
\centerline{\epsfbox{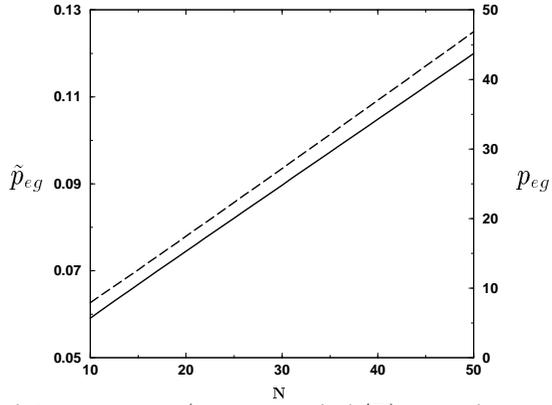}}
\caption {A comparison of $\tilde{p}_{ge}$ and $p_{ge}$ (in units
of $A/\Gamma$) as a function of $N$. The short interval $\tau$ is taken to
be one half of the bath correlation time.} 
\end{figure}

\begin{figure}
\epsfxsize 3 in
\centerline{\epsfbox{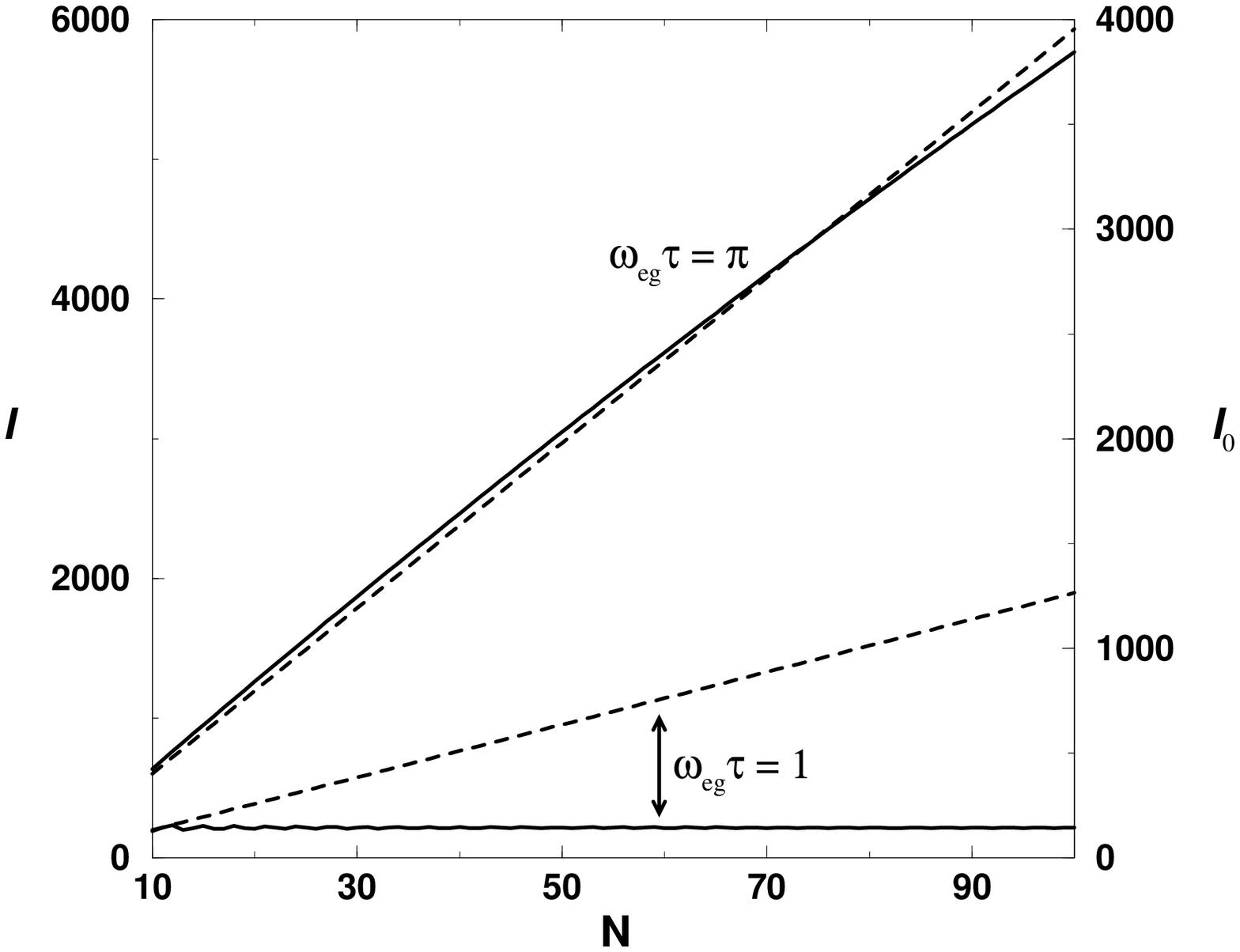}}
\caption {A comparison of I (Solid) (Eq.(19)) and $I_o$ (dashed) obtained
by dropping tan function in the integrand of Eq.(19) when we introduce
a cut-off at $x=1$. We show results for $\omega_{eg}
\tau = 1$ and $\pi$.}
\end{figure}

\begin{thebibliography}{00}
\bibitem{taylor} We have used for simplicity a Taylor series 
solution  of the Schr\"{o}dinger equation. Similar results are 
obtained  by  using (say) the full evolution operator or by
density matrix methods.
\end{thebibliography}
\end{document}